# A 40 Gbps Optical Transceiver for Particle Physics Experiments

B. Deng,[a] X. Huang,[b] H. Sun,[b,c,1] C.-P. Chao,[d] S.-W. Chen,[d] D. Gong,[b] S. Hou,[e] G. Huang,[c] C.-Y. Li,[d] C. Liu,[b] T. Liu,[b,2] Q. Sun,[f] J. Ye,[b] L. Zhang,[b,c,1] W. Zhang,[b,c,1]

[a] *Hubei Polytechnic University,*
 *Huangshi, Hubei 435003, China*

[b] *Southern Methodist University,*
 *Dallas, Texas 75275, USA*

[c] *Central China Normal University,*
 *Wuhan, Hubei 430079, China*

[d] *APAC Opto Electronics Inc.,*
 *Hsinchu, 303, Taiwan*

[e] *Academia Sinica,*
 *Nangang, Taipei 11529, Taiwan*

[f] *Fermi National Accelerator Laboratory,*
 *Batavia, IL 60510, USA*

 E-mail: tliu@smu.edu

ABSTRACT: We present the design and the test results of a quad-channel optical transceiver module (QTRx) possibly for future particle physics experiments. The transmitters of QTRx, each at 10 Gbps, are based on a Quad-channel VCSEL Diode array Driver (QLDD) and 1 × 4 VCSEL array. The receivers of QTRx, with data rates of 2.56 Gbps or 10 Gbps per channel, are based on a Quad-channel Trans-Impedance and limiting Amplifier (QTIA) and 1 × 4 photodiode array of GaAs or InGaAs. QTRx is 20 mm × 10 mm × 5 mm and couples to an MT fiber connector. Test results indicate that QTRx achieves the design goals with a power consumption of 124 mW per transmitter channel at 10 Gbps and 120 mW at 2.56 Gbps for the receiver channel with an on-chip charge pump. The sensitivities of QTIA are -17 dBm at 2.56 Gbps and -8 dBm at 10 Gbps, respectively. Further improvements with a gold-finger interface and a more compact optical lens are being designed.

KEYWORDS: Front-end electronics for detector readout; Radiation-hard electronics; Modular electronics.

---

[1] Visiting scholars at SMU and performed this work at SMU.
[2] Corresponding author.

# Contents

## 1. Introduction

With the increasing data volume in particle physics experiments, on-detector optical transceivers, such as VTRx [1], MTx/MTRx [2], VTRx+ [3], have been developed. These transceivers take advantage of commercial Vertical Cavity Surface Emitting Lasers (VCSELs) and Photodiodes (PDs). Studies have shown that the performances of VCSELs and PDs degrade with environmental parameters. The forward voltage of VCSELs increases with radiation and low temperature, resulting in insufficient voltage headroom [4]. The dark current and the junction capacitance of InGaAs photodiodes increase with radiation, whereas the responsivities of GaAs photodiodes decrease with radiation [5]. A Quad-channel Laser Diode Driver (QLDD) and Quad-channel Trans-Impedance and limiting Amplifier (QTIA) [6] have been developed to mitigate these performance degradations. In this paper, we present a Quad-channel optical transceiver module (QTRx) based on QLDD and QTIA. QTRx is a pure R&D project to gain experience in miniature optical modules. It is also a vehicle to test QLDD and QTIA. QTRx does not target any existing detector readout projects. The experience gained, as well as the actual module developed, might find usefulness and applications in future particle physics experiments, though, should occasion arise.

## 2. ASIC overview

QLDD is a derivative of the cpVLAD [7] with a larger chip area and robust implementation for production. QLDD implements an integrated charge pump and boosts the power supply of the output driver to enhance the headroom of the VCSEL driver. Figure 1(a) is the block diagram of QLDD. QLDD has four channels, each operating at 10 Gbps. Each channel consists of a limiting amplifier powered by 1.2 V and an output driver powered by the integrated charge pump. The charge pump in each channel can boost power voltages of 1.2 V and 2.5 V to a voltage up to 3.2 V. The negative feedback consists of an RC filter and an operational amplifier. The feedback can adjust the output voltage of the charge pump to keep the voltage difference between the output of the charge pump and the VCSEL anode constant, automatically adapting the forward voltage of the VCSEL during its lifetime [7]. The limiting amplifier and the output driver are AC coupled due to the differences in the common-mode voltages. QLDD is 2 mm × 2mm and fabricated in a 65 nm CMOS technology.

QTIA implements an integrated charge pump to provide a higher reverse bias voltage for the photodiode to overcome the issues mentioned in the introduction. The block diagram of QTIA is shown in figure 1(b). Each channel is composed of a bias structure to the PD, a TIA, a limiting amplifier, and an output driver. Four different bias structures to the PDs are used in different channels to study the mitigation options. The options include up bias (Channel 2), down bias (Channel 3), and dual bias (Channels 1 and 4). An up-bias structure uses a PMOS with a source degeneration to connect the cathode of the PD to the bias voltage (VPD), whereas a down-bias structure has an NMOS with a source degeneration from the anode of the PD to the ground [6]. In the dual-bias option, both up-bias and down-bias structures are implemented. In Channel 1, a three-stage charge pump is integrated to provide an internal high voltage over 5 V. The charge pump provides a potential solution for the voltage drop caused by the radiation-induced leakage-increase of the PD. Channel 4 uses an external high voltage to the PD as a reference of the integrated charge pump in Channel 1. QTIA is optimized for operating at 2.56 and 10 Gbps per channel. QTIA is 2 mm × 2 mm and fabricated in a 65 nm CMOS technology.

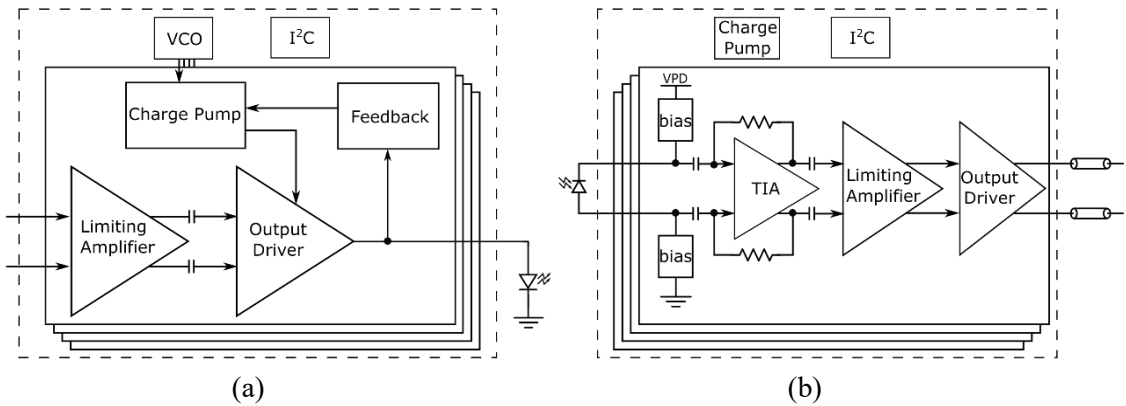

**Figure 1.** Block diagrams of QLDD (a) and QTIA (b).

## 3. Module design

The block diagram of QTRx is shown in figure 2. The transmitters of QTRx are based on QLDD and a 1 × 4 VCSEL array, and the receivers of QTRx are based on QTIA and a 1 × 4 photodiode array of either GaAs or InGaAs. Each transmitter or receiver channel can be disabled via Inter-Integrated Circuit (I$^2$C) to save power, so the number of transmitters and receivers can be reconfigured depending on different scenarios.

The differential electrical signals are AC coupled to QLDD and QTIA. The electrical interface is a 40-pin header (Hirose Electric, Part No. DF40C-40DP-0.4V(51)). The VCSEL and PD arrays are directly wire-bonded to QLDD and QTIA, respectively. QLDD and QTIA share the same I$^2$C bus with differing I$^2$C addresses. A lens (Orange Tek, Part No. OT-005) is utilized to couple the optical signals to a 12-fiber ribbon with an MT connector. Among the 12 fibers, the middle four fibers are unused, and each of the other eight fibers either transmits a 10 Gbps signal or receives a 2.56 or 10 Gbps signal. QLDD, QTIA, VCSEL array, PD array, and the bonding wires are covered and protected underneath the lens.

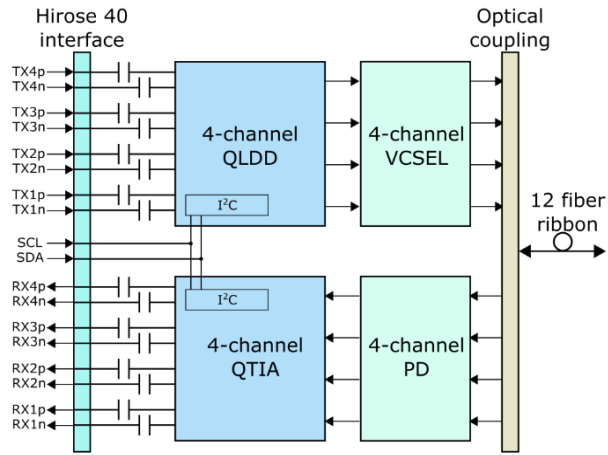

**Figure 2.** Block diagram of QTRx.

A photograph of a QTRx module mounted on a carrier board is shown in figure 3(a). The receptacle (Hirose Electric, Part No. DF40C-4.0-40DS-0.4V(51)) is mounted on the carrier board with a 4 mm stacking height. Eight pairs of 2.92 mm connectors are used for the differential electrical signals of QTRx. Two additional aluminum blocks are used to provide temporary mechanical support for QTRx on the carrier board. Figure 3(b) is a photograph of QTRx without fibers. All components of QTRx are located on one side of the module PCB, leaving the other side to be coupled to a cooling plate if needed. QTRx is 20 mm (L) × 10 mm (W) × 5 mm (H). The bonding wires, QLDD, and QTIA underneath the lens under the microscope are illustrated in figure 3(c). The slow-control signals and the power supplies are bonded to the top edge of QLDD or the bottom edge of QTIA. Further improvements in the module design will be discussed in Section 5.

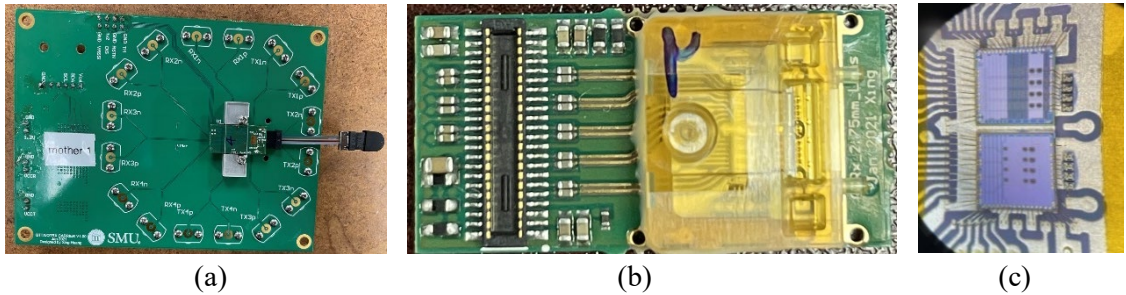

(a)  (b)  (c)

**Figure 3.** Photographs of a QTRx mounted on the motherboard (a), a QTRx (b), bonding wires of QTRx under a microscope (c).

## 4. Module tests

QTRx was characterized with the test board shown in figure 3(a). The block diagrams of QTRx test setups are presented in figure 4(a-b). To test transmitter channels, a pulse generator (Picosecond Lab, Model SDG 12070) generated a $2^7-1$ pseudo-random binary sequence (PRBS) differential signal to a transmitter channel of QTRx with 10-cm coaxial cables. An optical oscilloscope (Tektronix, Model TDS8000B) was used to test QTRx with a commercial 850 nm 25 Gbps VCSEL array (II-VI Laser Enterprise, Part No. APA450104002) through fibers. As for the receiver channels, an optical transmitter developed before [8] was used to transfer the electrical signal into an optical signal to the PDs of QTRx. In the test, 850 nm 28 Gbps GaAs PIN photodiodes (II-VI Laser Enterprise, Part No. APA1201040004) are used. The output of QTRx was tested by an electrical oscilloscope (Tektronix, Model DSA72004) and an error detector (Anritsu, Model MP1764C). The test setup photograph of the transmitter channels is shown in figure 4(c).

The input sensitivity of transmitter channels at 10 Gbps was tested by scanning the input voltage from 50 mV to 1.2 V. An input sensitivity of 80 mV was achieved. A 200 mV amplitude of the input signal was used to characterize the transmitter performance in all following tests. A typical eye diagram of a transmitter channel at 10 Gbps is shown in figure 5(a). The eye diagram passes the 10 Gbps eye mask test with a sufficient margin. The Average Optical Power (AOP) is 0.83 dBm, and the Optical Modulation Amplitude (OMA) is 428.3 μW. The extension ratio is 3 dB. The rise and fall times are around 37 ps.

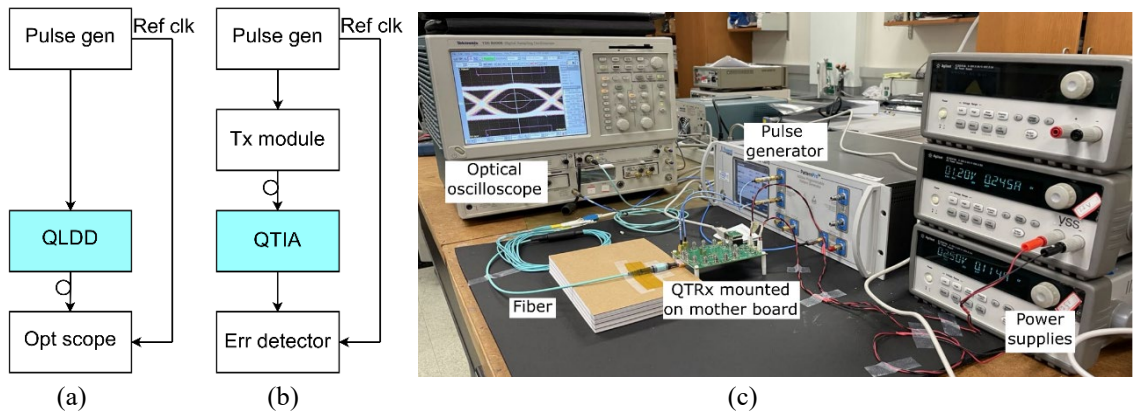

**Figure 4.** Test setup block diagrams of the transmitter (a) and receiver channels (b) and test setup photograph of transmitter channels (c).

The receiver channels were tested at 2.56 Gbps and 10 Gbps, respectively. The output eye diagram of a receiver channel with a -6 dBm input optical signal at 2.56 Gbps is shown in figure 5(b). The random jitter is 1.8 ps (RMS), and the deterministic jitter is 16.7 ps (P-P). The total jitter corresponding to the Bit Error Rate (BER) of $1\times10^{-12}$ is 38.5 ps (P-P). The rise/fall time at 2.56 Gbps is about 40 ps. The output eye diagram at 10 Gbps is shown in figure 5(c). The random jitter, deterministic jitter, and the total jitter are 2.4 ps (RMS), 22.7 ps (P-P), and 52.4 ps (P-P), respectively. The rise/fall times at 10 Gbps are about 50 ps.

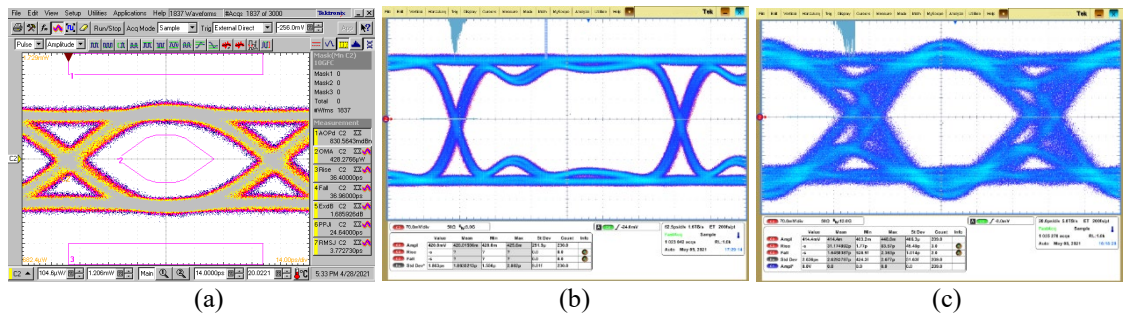

**Figure 5.** Eye diagrams of a transmitter channel at 10 Gbps (a), a receiver channel at 2.56 Gbps (b), and a receiver channel at 10 Gbps (c).

The sensitivity of receiver channels at the BER of $1\times10^{-12}$ was tested by scanning the input optical power. The BER versus the input optical power at 2.56 Gbps and 10 Gbps are shown in figures 6(a) and 6(b), respectively. In figure 6(a), CH1 and CH4 achieve a sensitivity of around -17 dBm, whereas CH2 and CH3 achieve around -14 dBm. The dual-bias structure provides a 3 dB improvement in comparison with the single-bias structures. The sensitivity of the receiver channels at 10 Gbps is about -8 dBm. The improvements induced by the charge pump will be verified in future irradiation tests.

The power consumptions of QTRx with on-chip charge pumps are 124 mW per transmitter channel at the modulation current of 7.5 mA and 120 mW per receiver channel. Compared with QTRx, VTRx+ consumes 92 mW per transmitter channel at the modulation current of 6 mA and 100 mW per receiver channel. The power consumption of QTRx is comparable with that of VTRx+ when the efficiency of charge pumps is taken into account. No significant crosstalk was observed in any channels of QTRx.

## 5. Further improvements

The electrical connector of the current prototype is fragile and is specified to have a maximum of 15 mating cycles. A new module with a more user-friendly gold finger interface (Samtec FireFly micro flyover system, Part No. UEC5-019-2-X-D-RA-1 and UCC8-010-1-X-S-1-A) has been brought forth to overcome the mechanical issues above. The new electrical interface provides natural mechanical support. The new module has a comparable dimension of 21 mm (L) × 10.5 mm (W) × 3.7 mm (H) in figure 7(a) with the same lens as shown in figure 3(b) and a much more compact profile of 16 mm (L) × 10.5 mm (W) × 2.3 mm (H) with another lens (Orange Tek, Part No. OT-012) in figure 7(b). The new module designs have been finalized and will be submitted for fabrication soon.

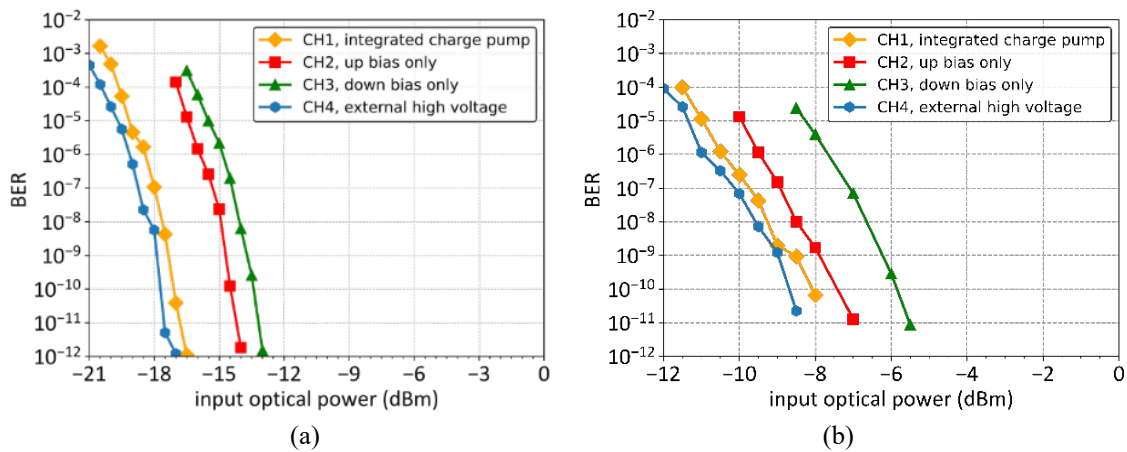

**Figure 6.** BER versus input optical power of the receiver channels at 2.56 Gbps (a) and 10 Gbps (b).

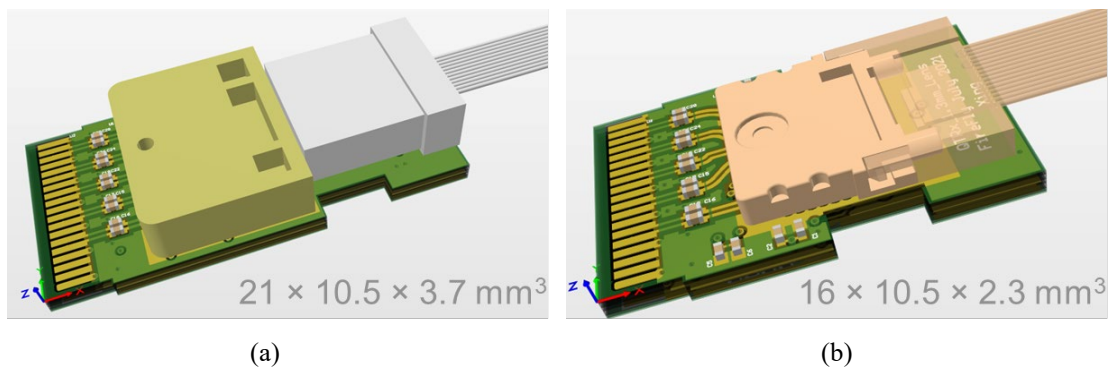

**Figure 7.** New prototype versions with a gold-finger interface (a) and a compact lens (b).

# 6. Conclusion

A prototype of a quad-channel optical transceiver module QTRx has been designed and tested, based on QLDD and QTIA. Each channel of QLDD can operate at 10 Gbps, whereas each channel of QTIA can operate at 2.56 Gbps or 10 Gbps. Charge pumps have been integrated into QLDD and QTIA to mitigate the performance degradations of VCSELs and PDs induced by radiation or low temperature. The prototype QTRx is 10 mm × 20 mm × 5 mm. The output eye diagrams of the transmitter channels pass the eye mask tests with a sufficient margin. The sensitivities of the receiver channels are -17 dBm at 2.56 Gbps and -8 dBm at 10 Gbps, respectively. The power consumptions of the transmitter and receiver channels with charge pumps are 124 mW/ch and 120 mW/ch, respectively. New prototypes with a gold finger electrical interface and a more compact lens have been designed and will be manufactured soon.

## Acknowledgments

This work is supported by SMU's Dedman Dean's Research Council Grant and the National Science Council in Taiwan.